\documentclass[12pt,letterpaper,]{article}
\usepackage{graphicx,epsfig,url,svg,fullpage}
\usepackage{caption}
\usepackage{float}
\usepackage{comment}
\usepackage{natbib}
\bibpunct{(}{)}{;}{a}{,}{;}

\usepackage{multicol}

\usepackage{fancyhdr}
\usepackage[headheight=30pt,headsep=1cm]{geometry}
\RequirePackage[tt=false, type1=true, sfdefault]{libertine}
\RequirePackage[libertine]{newtxmath}
\usepackage[caption=false]{subfig}

\title{TREC 2020 Podcasts Track Overview}

\author{
  Rosie Jones$^1$, Ben Carterette$^1$, Ann Clifton$^1$, Maria Eskevich$^2$, \\ 
  Gareth J. F. Jones$^3$, Jussi Karlgren$^1$, \\ Aasish Pappu$^1$, Sravana Reddy$^1$, Yongze Yu$^1$ \\ \\ 
  $^1$ Spotify \\ 
  $^2$ CLARIN ERIC \\
  $^3$ Dublin City University \\
  }

\date{}

\begin{document}
\maketitle
\noindent\rule{\textwidth}{0.2pt}     %%% DEPARTING FROM STANDARD ABSTRACT LAYOUT HERE
\small \noindent
\textbf{Abstract} ~~~ The Podcast Track is new  at the Text Retrieval Conference (TREC) in 2020.  The podcast track was designed to encourage  research into  podcasts in the information retrieval and NLP research   communities. The track consisted of two shared tasks: segment retrieval and summarization, both based on a dataset of over 100,000 podcast episodes (metadata, audio, and automatic transcripts) which was released concurrently with the track. The track generated considerable interest, attracted hundreds of new registrations to TREC and fifteen teams, mostly disjoint between search and summarization, made final submissions for assessment. Deep learning was the dominant experimental approach for both search experiments and summarization. This paper gives an overview of the tasks and the results of the participants' experiments.
The track will return to TREC 2021 with the same two tasks, incorporating slight modifications in response to participant feedback.

\noindent\rule{\textwidth}{0.2pt}      %%% DEPARTING FROM STANDARD ABSTRACT LAYOUT HERE

\thispagestyle{empty}   %%% TO BLOCK PAGE NUMBER ON FIRST PAGE

\begin{multicols}{2}

\section{Introduction}

Podcasts are a growing medium of recorded spoken audio. They are more diverse in style, content, format, and production type than previously studied speech formats, such as broadcast news \citep{GarafoloEtAl2000} or meeting transcripts \citep{renals2008}, and they encompass many more genres than typically studied in video research \citep{2006smeaton_a}.  
They come in many different formats and levels of formality --  news journalism or conversational chat,  fiction or non-fiction. Podcasts have a sharply growing share of listening consumption~\citep{infinitedial2020} and yet have been relatively understudied. The medium shows great potential to become a rich domain for research in information access and speech and language technologies (among other fields), with many potential opportunities to improve user engagement and consumption of podcast content.
The TREC Podcast Track which was launched in 2020 is intended to facilitate research in language technologies applied to podcasts.

\subsection{Data}

\begin{figure*}[hbp]
\centering
\subfloat[Transcript snippet]{
\includegraphics[width=\textwidth]{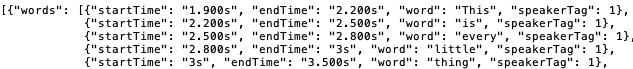}
}

\subfloat[Some of the accompanying metadata]{
\resizebox{0.9\textwidth}{!}{%
\begin{tabular}{r|p{4.5in}}
    Episode Name        & Mini: Eau de Thrift Store \\
    Episode Description & ELY gets to the bottom of a familiar aroma with 
                          cleaning expert Jolie Kerr. Guest: Jolie Kerr, 
                          of Ask a Clean Person. Thanks to listener Theresa. \\
    Publisher           & Gimlet \\
    RSS Link            & \url{https://feeds.megaphone.fm/elt-spot} \\
\end{tabular}}
}
\caption{Sample from an episode transcript and metadata}
\label{fig:transcript-example}
\end{figure*}

The data distributed by the track organisers consisted of just over 100,000 episodes of English-language podcasts. Each episode comes with full audio, a transcript which was automatically generated using Google's Speech-to-Text API as of early 2020, and a description and metadata provided by the podcast creator, along with the RSS feed content for the show. The data set is described in greater detail by \cite{clifton2020hundredthousand};  
an example is given in Figure~\ref{fig:transcript-example}.

\begin{table}[H]
\small
    \centering
    \begin{tabular}{c|c}
    Statistic Name &	Value    \\ 
    \hline
    Email list sign-ups & 	285 	 \\ 
In TREC slack channel  \#podcasts 2020 &	194	 \\
TREC podcasts registrations&	213	 \\
Signed data sharing agreement &	77	\\
Downloaded transcripts &	64 \\
Downloaded  audio &	18 \\
Participated in Search      & 7 \\
Participated in Summarization       & 8  \\
Participated in Both & 2  \\
\hline
\hline
    \end{tabular}
    \caption{Participation statistics}
    \label{tab:my_label}
\end{table}

\subsection{Participation} 
The Podcast Track attracted a great deal of attention with more than 200 registrations to participate. Most registrants did not submit experiments for assessment. After the submission deadline had passed, registrants were sent a questionnaire to establish what they found to be the biggest challenge when working on their experiment and their submission, and if they did not submit a result, what the most important challenge they found to stand in the way of submission. Participants were also asked to suggest how participation might be made easier for the coming year. The response rate was on the low side (10 responses) and the collated results indicate that the size of the data overwhelmed some participants. Suggestions for the coming year included organising a task for a subset of the data to enable new entrants to familiarise themselves with the problem space. 

\subsection{Tasks}

In 2020 the Podcast Track offered two tasks: (1) retrieval of fixed two-minute segments and (2) summarization of episodes. Both tasks were possible to complete on the automatic transcripts of episodes, rather than the audio data. The full audio data was provided, and teams were free to use it for their tasks (though only one team did do so, using the audio to improve the automatic transcription quality). The segment retrieval and summarization submissions were entirely based on textual input for all submitted experiments.

\section{Previous Work}

While there has been relatively little published work exploring information access technologies for podcasts, there is longstanding interest in spoken content retrieval in a range of other settings involving spoken content. 

\subsection{Spoken Document Retrieval}
The best known of work in spoken document retrieval is the TREC Spoken Document Retrieval Track which ran at TREC from 1997-2000 \citep{GarafoloEtAl2000}. The track focused on examining spoken document retrieval for broadcast news from radio and television sources of increasing size and complexity with each edition of the task. Task participants were provided with baseline transcripts of the spoken created using a then state-of-the-art automatic speech recognition (ASR) system and accurate or near-accurate manual transcripts of the content. The track began by using documents created by manually segmenting the news broadcasts into stories, but latterly began to explore automated identification of start points within unsegmented news broadcasts. The key findings were that for broadcasts, similar retrieval effectiveness could be achieved for errorful automatic speech recognition transcripts as for manual transcripts, through the appropriate use of external resources such as large contemporaneous news text archives.

A very different spoken retrieval task ran at the CLEF conference in the years 2005-2007 as the Cross-Language Speech Retrieval (CL-SR) task \citep{PecinaEtAl2008}. This focused on retrieval from a large archive of oral history  --- spontaneous conversations in the form of personal testimonies. Participants were provided with automatic speech recognition (ASR) transcripts of the spoken content, with a diverse set of associated metadata, manually and automatically assigned controlled vocabulary descriptors for concepts presented in each oral testimony, dates and locations associated with the content discussed, manually assigned person names, and expert hand-written segment summaries of the events discussed, together with a set of carefully designed search topics. The main task was to identify starting points for cohesive stories within the each conversational testimony interview where the ground-truth story boundaries were manually assigned by domain experts. The main findings of this task were that accurate automated location of topic start points is  challenging, and that, importantly, conversations of this type frequently fail to include mention of important entities within the dialogue. This means that search queries which include these entities often fail to match well with relevant content. This contrasts with search of broadcast news where such entities are mentioned very frequently to enable listeners to news updates can easily understand the events being described. Retrieval effectiveness was greatly improved by judicious use of the provided manual metadata, but it was recognised that such metadata will not be available for many spoken content archives.

Another spoken content retrieval task was offered at NTCIR from 2010-2016. This focused on search of Japanese language lectures and technical presentations. The first phase of the task focused only the retrieval of spoken content \citep{AkibaEtAl2013} while the second phase included the additional complexity of spoken queries \citep{AkibaEtAl2016}. As well as issues for automated transcription relating of the unstructured informal nature of the spoken delivery of this content, transcription of this content introduced challenges of transcription of specialised domain specific vocabulary items. Participants were provided with a set of search topics with a requirement to locate relevant content within the transcripts. A unique feature of this dataset was the very detailed fine-granularity labelling of relevant content for each search query within the transcripts. This meant that it was possible to do very detailed analysis of the ability of search methods to identify relevant content, including the relationship between search behaviour and the accuracy of the transcription of the query search terms within the transcripts.

A further study of spoken content search was the Rich Speech Retrieval and Search and Hyperlinking tasks at Mediaeval from 2011-2015 \citep{LarsonEtAl2011, EskevichEtAl2012, EskevichEtAl2015}. The primary search focus of this task was the identification of ``jump-in'' points in multimedia content based on the spoken soundtrack. In different years the task focused on different multimedia content collections. Initially the Blip10000 collection of crawled content from the {\tt blip.tv\/}\footnote{\url{https://en.wikipedia.org/wiki/Blip_(website)}} online platform of semi-professional user generated (SPUG) content \citep{Schmiedeke2013, EskevichEtAl2012} and later a collection of diverse broadcast television content provided by the BBC \citep{EskevichEtAl2015}. Participants were provided with state-of-the-art ASR transcripts of the content archives and carefully developed search queries. Tasks included known-item and ad hoc search, with relevance assessment using crowdsourcing methods. As well as confirming earlier findings in terms of automated location of useful jump-in points, there was significant focus in these tasks on how submissions should be comparatively evaluated. In particular, the trade-off between ranking of retrieved items containing relevant content and the accuracy of the identification jump-in points in retrieved items.  

As well as these benchmark tasks, another relevant study in spoken content retrieval using the AMI corpus \citep{renals2008} is reported in \cite{eskevich2014csl} which gives a detailed examination of the differences in the ranking of retrieved items between manual and automated transcripts arising from ASR errors. A more complete overview of research in spoken content retrieval from its beginnings in the early 1990s to today can be found in \cite{jones2019}. While none of this existing work focuses on podcast search, the various content archives used raise many of the same issues that can be observed in podcasts in terms of content diversity, use of domain specific vocabularies, and probable issues relating to absence of entity mentions in conversational podcasts. 

\subsection{Summarization}
% add a small section on summarization?
While there is a great deal of work on summarizing text in the news domain (eg~\cite{mihalcea2004textrank}), there is much less existing work on summarization of spoken content. One study relevant to the Podcast Track is that of \cite{SpinaTCS17}. This work focuses on the creation of 
query biased audio summaries of podcasts. A crowdsourced experiment demonstrated that highly noisy automatically generated transcripts of spoken documents are effective sources of document summaries to support users in making relevance judgements for a query. Particularly notable was the finding that summaries generated using ASR transcripts were comparable in terms of usability to summaries generated using error-free manual transcripts.

\subsection{Podcast Information Access}
\citet{Besser:2008:UserGoalsPodcastSearch} argues that the underlying goals of podcast search may be similar to those for blog search, as podcast can be viewed as audio blogs. 
In \cite{DBLP:journals/jasis/TsagkiasLR10}, the general appeal of podcast feeds/shows is predicted from various features. The authors identify as important factors of whether a user subscribes to a podcast feed: whether the feed has a logo, length of the description, keyword count, episode length, author count, and feed period.
\cite{yang2019more}  showed they could use acoustic features
to predict seriousness and energy of podcasts,  as well as popularity. Acoustic features take advantage of a unique aspect of podcasts, and can be used as part of a multimodal approach to podcast information access, which we hope to see more of in the track in future years.

\section{Segment Retrieval Task}
\subsection{Definition}
The retrieval task was defined as the problem of finding relevant segments from the episodes for a set of search queries which were provided in traditional TREC topic format. The provided transcripts have word-level time-stamps on a granularity of 0.1s which allows retrieval systems to index the contents by time offsets. A segment was defined to be a two-minute chunk starting on the minute; e.g. [0.0-119.9] seconds, [60-199.9] seconds, [120-139.9] seconds, etc. Segments overlap each other by one minute - any segment except for the first and last segment is covered by the preceding and following segments. The rationale for creating overlapping segments is to account for the case where a phrase or sentence is split across segment boundaries.
This creates 3.4M segments in total from the document collection with an average word count of $340 \pm 70$ per segment.
Topics consist of a topic number, keyword query, a type label, and a description of the user’s information need. Eight topics were given at the outset for the participants to practice on, and 50 topics were released as the test task. Topics were formulated in three \textit{types}: topical, re-finding, and known item. Example topics are given in Figure~\ref{queryexample}.

\begin{figure*}
    \centering
    \begin{verbatim}
<topic>
<num>34</num>
<query>halloween stories and chat</query>
<type>topical</type>
<description>I love Halloween and I want to hear stories and conversations 
             about things people have done to celebrate it.  I am not looking 
             for information about the history of Halloween or generalities 
             about how it is celebrated, I want specific stories from 
             individuals.
</description>
</topic>

<topic>
<num>45</num>
<query>drafting tight ends</query>
<type>refinding</type>
<description>I heard a podcast about strategies for drafting tight ends in 
             football.  I’d like to find it again.
</description>
</topic>

<topic>
<num>58</num>
<query>sam bush interview</query>
<type>known item</type>
<description>A bluegrass magazine I read mentioned a podcast interview with 
             Sam Bush.  I’d like to hear it.
</description>
</topic>


\end{verbatim}
    \caption{Example search topics}
    \label{queryexample}
\end{figure*}

\subsection{Submissions}
7 participants submitted 24 experiments for the retrieval task. All runs were `automatic', i.e, without human intervention; almost all runs were based on the Query Description field, i.e. the more verbose exposition of information need as shown in Figure~\ref{queryexample}. For training data, many participants used pretrained transfer learning models, some used language technologies and knowledge-based models, and some used only data from the set as shown in table~\ref{tab:retrtech}. Only one experiment made use of the audio data to produce and use a different transcript than the provided one. 

\begin{table*}[]
    \centering
\begin{tabular}{ll|ccll}
\textbf{Participant}     &       \textbf{run id}            &  \textbf{field}     &    \textbf{transfer}  &  \textbf{data}               &  \textbf{IR}          \\
                &                         &         &    \textbf{learning}  &  \textbf{processing}            &            \\
\hline                                              
Dublin City U	& 	dcu1		  &   D     &              &  SpaCy                  & QE from WordNet       \\ 
 		& 	dcu2		  &   D     &              &  SpaCy                  & QE from Descriptions         \\
 		& 	dcu3		  &   D     &              &  Spacy                  & QE, auto RF         \\
 		& 	dcu4		  &   D     &              &  Spacy                  &  QE from web text        \\
 		& 	dcu5		  &   D     &              &  Spacy                  &  Combination 1-4        \\
\hline                                              
LRG        & LRGREtvrs-r\_1  &   D     &  \checkmark  &                         & XLNet;Regression        \\
           & LRGREtvrs-r\_2  &   D     &  \checkmark  &                         & XLNet;Regression+Concat     \\  
           & LRGREtvrs-r\_3  &   D     &  \checkmark  &                         & XLNet;Similarity       \\
\hline                                              
U Maryland	& 	UMD\_IR\_run1	  &   D     &  \checkmark  &                         &  Indri            \\ 
		& 	UMD\_IR\_run2	  &   D     &              &                         &  Indri            \\ 
		& 	UMD\_IR\_run3	  &   D     &  \checkmark  & stemming                &  Combination + Rerank           \\
		& 	           	  &         &              & word2vec                &             \\
                & 	UMD\_ID\_run4	  &   D     &  \checkmark  & stemming                &  rerank + Combination \\
                & 	           	  &         &              &  word2vec               &   \\
                & 	UMD\_IR\_run5	  &   D     &  \checkmark  & stemming                &  Combination of 1-4       \\
\hline                                              
U Texas Dallas	& 	UTDThesis\_Run1	  &   D     &  \checkmark  &  fuzzy match            & Lucene            \\ 
\hline                                              
Johns Hopkins   &	hltcoe1		  &   Q     &              &  5-gram                  &  Rocchio RF          \\
HLT COE 	&	hltcoe2		  &   Q     &              &                          &  Rocchio RF            \\
		&	hltcoe3		  &   Q     &              &                          &  no RF            \\
		&	hltcoe4		  &   D     &              &                          &  Rocchio RF           \\
		&	hltcoe5		  &   Q     &              &  transcript              &  Rocchio RF   \\
		&			  &         &              &   4-gram                 &     \\
\hline                                              
U Oklahoma	&	oudalab1	  &   D     & \checkmark   &  SpaCy                   & BM25; Faiss; finetuned on SQuAD   \\ % bm25
\hline                                              
Spotify		&	BERT-DESC-S	  &   D     & \checkmark   &                          & rerank 50;  \\
		& & & & & finetuned on other topics  \\
		&	BERT-DESC-Q	  &   D     & \checkmark   &                          & rerank 50; \\
		& & & & & finetuned on automatic topics  \\ 
		&	BERT-DESC-TD	  &   D     & \checkmark   &                          & rerank 50;  \\
		& & & & & finetuned on synthetic data  \\
\hline                                              
\hline                                              
baseline        &	BM25	          &   Q     &              &                          & BM25             \\
                &	QL	          &   Q     &              &                          & query likelihood             \\
                &	RERANK-QUERY	  &   Q     & \checkmark   &                          & rerank 50            \\
                &	RERANK-DESC	  &   D     & \checkmark   &                          & rerank 50             \\
\end{tabular}
    \caption{Technologies employed for the retrieval task}
    \label{tab:retrtech}
\end{table*}

\subsection{Evaluation}
Two-minute length segments were judged by NIST assessors for their relevance to the topic description. NIST assessors had access to both the ASR transcript (including text before and after the text of the two-minute segment, which can be used as context) as well as the corresponding audio segment. Assessments were made on the PEGFB graded scale (Perfect, Excellent, Good, Fair, Bad) as approximately follows:
\begin{description}

\item[Perfect (4):] this grade is used only for ``known item" and ``refinding" topic types. It reflects the segment that is the earliest entry point into the one episode that the user is seeking.

\item[Excellent (3):] the segment conveys highly relevant information, is an ideal entry point for a human listener, and is fully on topic. An example would be a segment that begins at or very close to the start of a discussion on the topic, immediately signaling relevance and context to the user.

\item[Good (2):] the segment conveys highly-to-somewhat relevant information, is a good entry point for a human listener, and is fully to mostly on topic. An example would be a segment that is a few minutes “off” in terms of position, so that while it is relevant to the user’s information need, they might have preferred to start two minutes earlier or later.

\item[Fair (1):] the segment conveys somewhat relevant information, but is a sub-par entry point for a human listener and may not be fully on topic. Examples would be segments that switch from non-relevant to relevant (so that the listener is not able to immediately understand the relevance of the segment), segments that start well into a discussion without providing enough context for understanding, etc.
\item[Bad (0):] the segment is not relevant.
\end{description}

\begin{figure}[H]
    \centering
    \includegraphics[width=3in]{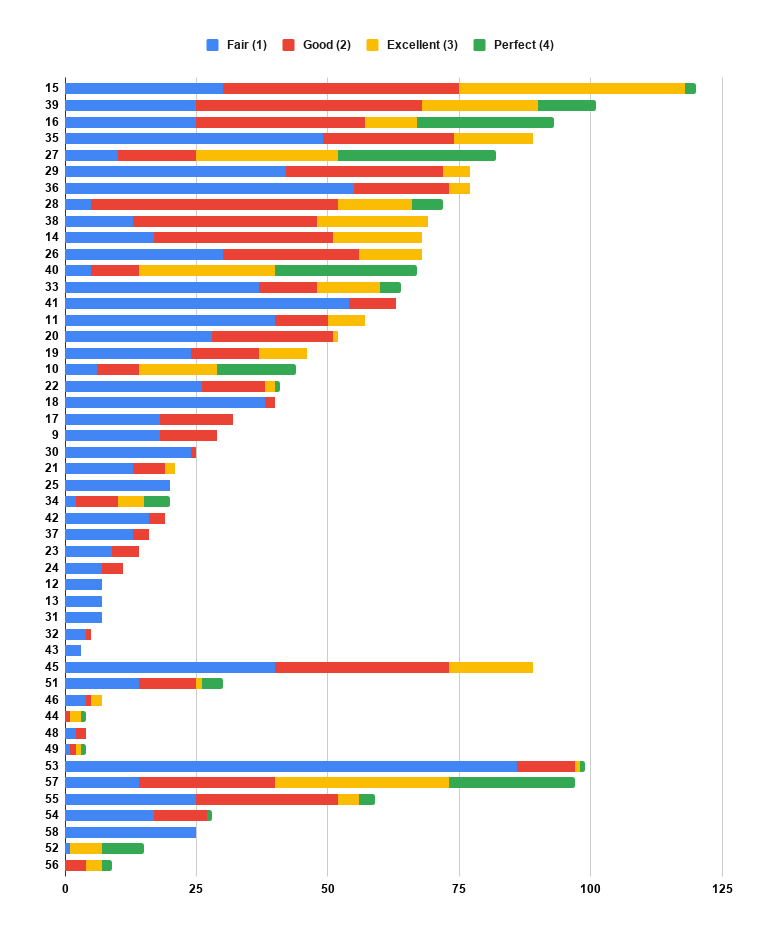}
    \caption{Number of relevant segments of different type per topic, ranged by the number of relevant episodes per three topics categories (topical (15-43), refinding (45-49), known items (53-56).}
    \label{fig:segment-qrel-overview}
\end{figure}

Figure \ref{fig:segment-qrel-overview} shows the number of relevant segments of different type per topic. The results are ranged into three groups based on the topic types: topical (15-43), refinding (45-49), known items (53-56). This demonstrates that all topics had some relevant segments retrieved by participants and assessed by assessors.   

The primary metric for evaluation is mean nDCG, with normalization based on an ideal ranking of all relevant segments.  Note that a single episode may contribute one or more relevant segments, some of which may be overlapping, but these are treated as independent items for the purpose of nDCG computation.

\subsection{Search Baselines}

Podcast search could be implemented without the full episode transcripts if the titles and creator-provided descriptions provide enough information for search and indexing. As a first baseline, we compared document level retrieval of transcripts to document level retrieval based on titles and creator-provided descriptions.
Table~\ref{tab:transcript} shows how using transcripts yields vastly higher scores, compared to using titles or descriptions, episode-level or episode and show-level combined. Even so, adding titles and descriptions to the transcripts improves results somewhat. In all subsequent experiments, the baseline use only transcripts, and do not use show or episode title or descriptions.

\begin{table*}[]
    \centering
    \begin{tabular}{l|ccc}
				              & nDCG	& nDCG at 30 & precision at 10 \\
    \hline
Episode Title			      & 0.22	& 0.19	& 0.12  \\
Episode Description		      & 0.32	& 0.27	& 0.17 \\
Episode Title and Description & 0.36	& 0.30	& 0.19 \\
Episode Title and Description & 		& 	& \\
~~ with Show Title and Description   
                              & 0.37	& 0.30	& 0.20 \\
Transcript Text        		  & 0.58	& 0.46	& 0.41 \\
Transcript Text 		      & 		& 	& \\
~~with Episode Title and Description 
                              & 0.61	& 0.49	& 0.43 \\

    \end{tabular}
    \caption{The contribution of transcripts compared to title search on search results}
    \label{tab:transcript}
\end{table*}

Four baseline segment retrieval runs on transcripts are included, using both standard information retrieval methods as well as re-ranking models using BERT~\citep{devlin-etal-2019-bert} to represent segment content.
\begin{enumerate}
    \item BM25: Standard information retrieval algorithm developed for the Okapi system\footnote{Implemented using the Pyserini package, \url{ https://github.com/castorini/pyserini} --  a Python front end to the Anserini open-source information retrieval toolkit (\cite{yang2017anserini})}; the query field of the topic was used for search terms, and up to 1000 segments are returned for each topic. 
    \item QL (Query Likelihood): Standard information retrieval algorithm\footnotemark[2]
    \item RERANK-QUERY: A BERT re-ranking model pre-trained on MS MARCO passage retrieval data \citep{nogueira2019passage} without further parameter tuning; the query of the topic was used as the input to the re-ranking model; the re-ranking scores of top-50 segments from BM25 were calculated and submitted per topic. 
    \item RERANK-DESC: Same as RERANK-QUERY except that the description of the topic was used as the input in re-ranking model. 
\end{enumerate}

\subsection{Search Results}
\begin{table*}[]
    \centering
    \begin{tabular}{l|ccc}
                           & nDCG   & nDCG at 30 & precision at 10 \\
    \hline
    \hline
UMD\_IR\_run3	&	0.67	&	0.52	&	0.60	\\
UMD\_ID\_run4	&	0.66	&	0.49	&	0.56	\\
UMD\_IR\_run1	&	0.62	&	0.45	&	0.53	\\
UMD\_IR\_run5	&	0.65	&	0.50	&	0.58	\\
UMD\_IR\_run2	&	0.59	&	0.42	&	0.51	\\
run\_dcu5	&	0.59	&	0.43	&	0.54	\\
run\_dcu4	&	0.58	&	0.42	&	0.54	\\
run\_dcu1	&	0.57	&	0.42	&	0.50	\\
run\_dcu3	&	0.57	&	0.42	&	0.50	\\
run\_dcu2	&	0.55	&	0.40	&	0.48	\\
LRGREtvs-r\_2 * & 0.54  &   0.40    &   0.48    \\
LRGREtvs-r\_1 * & 0.54  &   0.40    &   0.47    \\
hltcoe4		&	0.51	&	0.43	&	0.54	\\
LRGREtvs-r\_3 * & 0.50  &   0.32    &   0.41    \\
hltcoe3		&	0.50	&	0.35	&	0.43	\\
hltcoe2		&	0.47	&	0.38	&	0.45	\\
hltcoe1		&	0.45	&	0.33	&	0.38	\\
BERT-DESC-S	&	0.43	&	0.47	&	0.57	\\
BERT-DESC-TD	&	0.43	&	0.47	&	0.56	\\
BERT-DESC-Q	&	0.41	&	0.45	&	0.53	\\
hltcoe5		&	0.38	&	0.30	&	0.37	\\
UTDThesis\_Run1	&	0.34	&	0.34	&	0.43	\\
oudalab1	&	0.00	&	0.01	&	0.01	\\
    \hline
Baseline BM25              &   0.52     &   0.40    &   0.49    \\
Baseline QL                &   0.52     &   0.40    &   0.48    \\
Baseline RERANK-DESC       &   0.43     &   0.48    &   0.57    \\
Baseline RERANK-QUERY      &   0.43     &   0.47    &   0.56    \\
    \hline
    \end{tabular}
    \caption{Overview of results from submitted segment retrieval experiments. 
    Post-assessment rerun submissions are marked with an asterisk.
    }
    \label{tab:retrieval-results}
\end{table*}

Table~\ref{tab:retrieval-results} gives an overview of the scores for the submitted experiments. Scoring only the top 30 items or the top 10 items of the list promotes some reranking approaches to the top of the list, illustrating the effect of use case-motivated evaluation metrics on system comparison. One participant resubmitted results after the assessment, to redress the effects of a processing mishap, and those results are marked in the table with an asterisk.

\section{Summarization Task}
\subsection{Definition}
Given a podcast episode, its audio, and transcription, the task is to return a short text snippet which accurately conveys the content of the podcast. Returned summaries should be grammatical, standalone utterances of significantly shorter length than the input episode description, short enough to be quickly read on a smartphone screen.

No ground truth summaries are provided; the closest proxies are the show and episode descriptions provided by the podcast creators. We observe that these descriptions vary widely in scope, and are not always intended to act as summaries of the episode content, reflecting the different genres represented in the sample and the different intentions of the creators for the descriptions. We filtered the descriptions to establish a subset that is more appropriate as a ground truth set compared to full set of descriptions. The filtering was done with three heuristics shown in Table \ref{tab:brassFilters}. These filters overlap to some extent, and remove about a third of the entire set; the remaining 66,245 descriptions we call the {\em Brass Set}. 

\begin{table*}[]
\small
    \centering
    \begin{tabular}{c|l|r}
    filter          &   criteria                            & items affected \\
\hline
       Length       &  very long ($> 750$ characters) or    & $24,033$ ($23\%$) \\
                    &  very short ($< 20$ characters)       &  \\ 
     \hline
Similarity to       &  $> 50\%$ lexical overlap             &  $15,375$ ($15\%$) \\
other descriptions  &   with other episode descriptions     &  \\ 
 \hline
Similarity to       & $> 40\%$  lexical overlap             & $9,444$ ($9\%$) \\ 
show description    & with own show description             &   \\
    \end{tabular}
    \caption{Filters to remove `less descriptive' episode descriptions, to form the {\em brass subcorpus}. }
    \label{tab:brassFilters}
\end{table*}

\subsection{Submissions}
8 participants submitted 22 experiments for the summarization task (Table~\ref{tab:sumtech}). All experiments used some form of deep learning model, and while some used extractive filtering of material from the transcripts as a step in their processes, all were based on abstractive techniques. No participant used the audio data for summary generation. 

\begin{table*}[]
    \centering
\begin{tabular}{ll|cl}
  \textbf{Participant}     &       \textbf{run id}               &  \textbf{type}  & \textbf{method} \\
  \hline                                       
U New Hampshire    &   unhtrema1             &     O    & GAN, LSTM, 3 sentences, long chunks \\
                    &    unhtrema2             &     O    & GAN, LSTM, 10 sentences, long chunks \\
                    &    unhtrema3             &     O    & GAN, LSTM, 20 sentences, short chunks \\
                    &    unhtrema4             &     O    & GAN, LSTM, 10 sentences, short chunks \\
U Central Florida  &   UCF\_NLP1             &     A    & BART \\
                   &   UCF\_NLP2             &     A    & BART, RoBERTa \\
U Texas Dallas	   & 	UTDThesis\_Run1      &     A    & T5, fine tuned on brass set \\
 & & & $+$ Dialogue Action Tokens \\
U Glasgow          &  2306987O\_abs\_run1    &     A    &  T5, fine tuned on description \\
                   &  2306987O\_extabs\_run2 &     O    &  15 sentence input, T5 \\
                   &  2306987O\_extabs\_run3 &     O    &  Extractive filtering, SpanBert \\
U Cambridge        & cued\_speechUniv1       &     A    &  BART, sentence filtering, 9 model ensemble \\
                   & cued\_speechUniv2       &     A    &  BART, sentence filtering, 3 model ensemble \\
                   & cued\_speechUniv3       &     A    &  BART, Fine tuned on transcript \\
                   & cued\_speechUniv4       &     A    &  BART, sentence filtering, non-ensemble \\
Uppsala U          & hk\_uu\_podcast1        &     A    &  BART, Longformer, 3 epochs \\
          Spotify  & categoryaware1          &     A    &  BART, Fine tuned on start of transcript  \\
 & & & $+$ podcast category; 1 epoch \\
                   & categoryaware2          &     A    &  BART, Fine tuned on start of transcript  \\
 & & & $+$ podcast category; 2 epochs \\
                   & coarse2fine             &     A    &  BART, Fine tuned on TextRank center of transcript;  \\
 & & & 2 epochs \\
U Delaware         & udel\_wang\_zheng1      &     A    & Start of transcript, BART \\
                   & udel\_wang\_zheng2      &     A    & Select sentences by LDA, BART \\
                   & udel\_wang\_zheng3      &     A    & Select sentences by ROUGE, BART \\
                   & udel\_wang\_zheng4      &     A    & Ensemble of 1-3 \\
\hline                                              
\hline                                              
Baseline           &   bartcnn               &     A    &  BART, No fine tuning \\
                   &   bartpodcasts          &     A    &  BART, Fine tuned on start of transcript \\
                   &   onemin                &     E    &  1 minute of transcript \\
                   &   textranksegments      &     E    &  TextRank, 50 wd segments \\
                   &   textranksentences     &     E    &  TextRank, sentence split \\
\end{tabular}
    \caption{Technologies employed for the summarization task}
    \label{tab:sumtech}
\end{table*}

\subsection{Evaluation} 
The summary labels and scores for the participating systems are created for evaluation sets in two ways.

 \subsubsection{Manual Assessments and Scoring}
Summaries are judged on a four-step scale intended to model how well a listener is able to make a decision whether to listen to a podcast or not, conveying a gist of what the user should expect to hear listening to the podcast. The assessment scale used by the NIST assessors is the EGFB scale, as per the following instructions:

\begin{description}
\item[Excellent:] the summary accurately conveys all the most important attributes of the episode, which could include topical content, genre, and participants. In addition to giving an accurate representation of the content, it contains almost no redundant material which is not needed when deciding whether to listen. It is also coherent, comprehensible, and has no grammatical errors.

\item[Good:]  the summary conveys most of the most important attributes and gives the reader a reasonable sense of what the episode contains with little redundant material which is not needed when deciding whether to listen. Occasional grammatical or coherence errors are acceptable. 

\item[Fair:] the summary conveys some attributes of the content but gives the reader an imperfect or incomplete sense of what the episode contains. It may contain redundant material which is not needed when deciding whether to listen and may contain repetitions or broken sentences. 

\item[Bad:] the summary does not convey any of the most important content items of the episode or gives the reader an incorrect or incomprehensible sense of what the episode contains. It may contain a large amount of redundant information that is not needed when deciding whether to listen to the episode. 
\end{description}

 NIST assessors evaluated 180 of the automatically-generated summaries produced by participants using the EGFB scale. These assessments are converted into a numerical score by a weighting scheme tested for being able separate the baseline systems applied to the Brass set. Weights of 4-2-1-0 for EGFB turned out to be simple and effective in this respect. 

In addition to the EGFB assessments, we created a set of boolean attributes that a desirable podcast summary might contain. The primary evaluation metric is the EGFB score; the answers to these attributes are merely an informative signal for participants, and may be useful in devising automated summarization metrics in the future.
The attributes are defined from a small-scale survey of podcast listeners, and are listed below.

\begin{enumerate}
    \item {\bf names:}
Does the summary include names of the main people (hosts, guests, characters) involved or mentioned in the podcast?
\item {\bf bio:} Does the summary give any additional information about the people mentioned (such as their job titles, biographies, personal background, etc)?
\item {\bf topics:}  Does the summary include the main topic(s) of the podcast?
\item {\bf format:} Does the summary tell you anything about the format of the podcast; e.g. whether it's an interview, whether it's a chat between friends, a monologue, etc?
\item {\bf title-context:} Does the summary give you more context on the title of the podcast?
\item {\bf redundant:} Does the summary contain redundant information?
\item {\bf english:} Is the summary written in good English?
\item {\bf sentence:} Are the start and end of the summary good sentence and paragraph start and end points?
\end{enumerate}

 \subsubsection{ROUGE against Creator Descriptions} Each of the test set episodes has a creator-provided description. This description is used as a reference (in the absence of ground truth summaries), and a ROUGE-L \cite{lin2004rouge} score against the description is computed. ROUGE-L computes overlap of substrings up to the length of the longest common subsequence. Note that these creator-provided descriptions are of varying quality: of the 179-sized subset of descriptions assessed by NIST, we find that only 71, between a third and half, are of Good or Excellent quality.
 
We provided a version of the episode descriptions processed by a BERT-based sentence classifier that was trained from a small set of manually annotated examples to identify and remove extraneous content such as boilerplate, ads, promotions, and show notes that do not directly summarize or describe the episode~\citep{podcastaddetect}. `Cleaned' descriptions with such extraneous content removed were produced, and
NIST asessors judged the cleaned descriptions as well as the original descriptions for summary quality. 

ROUGE scores were computed against the original episode descriptions, but may be computed against the `cleaned' descriptions as well.

\subsection{Summarization Baselines}
Five baseline summarization runs are included. We aimed to include a representation of abstractive as well as extractive models.

\begin{enumerate}
    \item onemin: Transcript text for the first one minute of the episode.
    \item bartcnn: A BART \citep{lewis-etal-2020-bart} seq2seq model pre-trained on the CNN/Daily Mail corpus for news summarization \footnote{https://huggingface.co/facebook/bart-large-cnn}
    \item bartpodcasts: The bartcnn model above, fine-tuned on the full 100k  episodes in the dataset, excluding episodes with very short (fewer than 10 characters) or very long descriptions (over 1300 characters). Episodes with descriptions that were highly similar\footnote{We represented each description by a normalized vector of TF-IDF values, and computed similarity as the cosine similarity between the vector representations.} to other descriptions in the same show, or to the show description itself, were also ignored. The descriptions were also processed through a model to detect and remove ads, promotions, and show notes such as links to transcripts. 
    \item textranksegments: We chunked the transcript into one-minute chunks, and applied the TextRank algorithm \citep{mihalcea2004textrank}, with word overlap as the similarity metric, to find the most `central' one-minute segment.
    \item textranksentences: The same process as above, except that we chunked the transcript into sentences using SpaCy\footnote{https://spacy.io} and extracted the two most central sentences.
\end{enumerate}

\subsection{Summarization Results} 

 179 episodes were scored for EGFB quality and the boolean attributes by NIST assessors for the 22 submitted experiments and the 5 baselines. The submitted experiments were also scored automatically using ROUGE-L against the creator-provided descriptions for all the 1024 test episodes. Table~\ref{tab:manualsummarisation-results} shows both the manual assessment scores as well as the automatic evaluations.

\begin{figure}[H]
    \centering
    \includegraphics[width=3in]{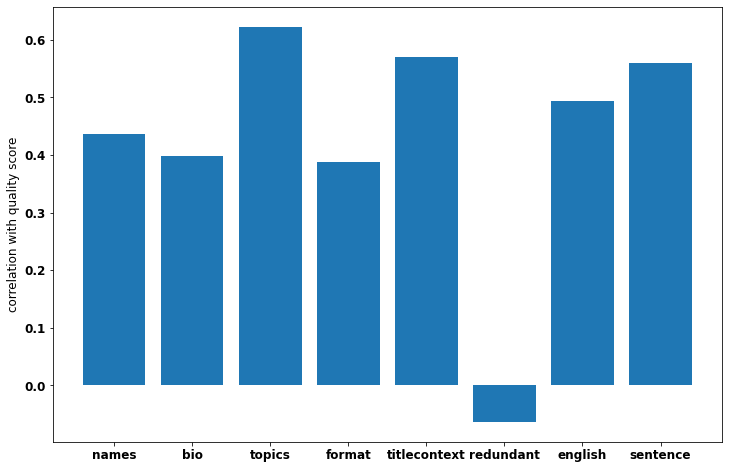}
    \caption{Pearson correlation of attributes with the aggregate EGFB quality score across all submitted baseline runs.}
    \label{fig:nuggets}
\end{figure}

All attributes were found to be significantly correlated with the aggregate quality score (Figure \ref{fig:nuggets}), to different degrees with `Does the summary include the main topic(s) of the podcast?' being the most correlated. Future work might investigate these attribute values across all submitted systems towards gaining a concrete understanding of what makes a good podcast summaries.

\begin{figure}[H]
    \centering
    \includegraphics[width=3in]{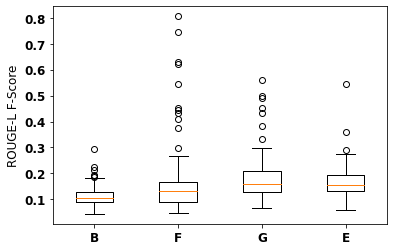}
    \caption{Distribution of ROUGE-L F-Score for each manually assessed label across all submitted baseline runs.}
    \label{fig:rouge}
\end{figure}

The ROUGE-L F-score is found to be weakly but significantly correlated with the aggregate EGFB quality score (Pearson correlation 0.28). As Figure \ref{fig:rouge} shows, while summaries rated E and G do have higher median ROUGE scores than those rated F and B, the variation is tremendously large, especially for summaries rated F, raising the question of whether ROUGE against creator descriptions is a sufficiently reliable metric for this task.

The episodes in the test set were variously challenging. Some very topical podcasts with a clear statement of purpose or a concise topical heading are comparatively easy to summarise, if that statement was identified in the episode transcript or even in the episode description: {\em ``Welcome to my podcast! Let us talk and learn about God's word, life-purpose, values, and faith!''}, or {\em ``On this day in 1826, 15-year-old Ellen Turner was abducted in a forced marriage plot intended to swindle her family out of their fortune.''} Episodes with a broad range of covered topics (such as the hosts' opinions on various books, movies and video games and their experiences from working in comedy), and episodes that are not about topics but rather, are sleep aids or avant-garde performance pieces, proved challenging for most systems.

\begin{table*}[]
    \centering
    \begin{tabular}{l|ccc|cc}
experiment & aggregate  & \#E & \#E,G & ROUGE-L & ROUGE-L \\	
& EGFB score & & & recall & precision \\
    \hline
    \hline
cued\_speechUniv2 & 2.04 & 47 & 105 & 	0.224		& 	0.235\\
cued\_speechUniv1 & 1.98 & 49 & 104 & 	0.226		& 	0.232 \\
cued\_speechUniv4 & 1.94 & 46 & 101 & 	0.204		& 	0.231	 \\
UCF\_NLP2 & 1.81 & 45 & 92 & 	0.224		& 	0.256 \\
cued\_speechUniv3 & 1.78 & 39 & 90 & 	0.205		& 	0.220 \\
hk\_uu\_podcast1 & 1.74 & 35 & 89 & 	0.190		& 	0.265	 \\
UCF\_NLP1 & 1.64 & 34 & 79 & 	0.220		& 	0.267\\
categoryaware2 & 1.58 & 32 & 71 & 	0.199		& 	0.257	\\
categoryaware1 & 1.51 & 26 & 75 & 	0.208		& 	0.227\\
coarse2fine & 1.3 & 18 & 57 & 	0.187		& 	0.158\\
udel\_wang\_zheng1 & 1.19 & 13 & 52& 	0.161		& 	0.239 \\
udel\_wang\_zheng4 & 1.16 & 14 & 53& 	0.168		& 	0.202 \\
udel\_wang\_zheng3 & 1.08 & 10 & 44 & 	0.160		& 	0.208\\
2306987O\_abs\_run1 & 1.00 & 12 & 39& 	0.156		& 	0.208 \\
2306987O\_extabs\_run2 & 0.99 & 13 & 42& 	0.167		& 	0.237 \\
2306987O\_extabs\_run3 & 0.80 & 8 & 22& 	0.147		& 	0.220 \\
udel\_wang\_zheng2 & 0.76 & 7 & 28 & 	0.139		& 	0.184\\
UTDThesis1 & 0.43 & 1 & 11	& 	0.129		& 	0.172\\
unhtrema4 & 0.04 & 1 & 1	& 	0.180		& 	0.069 \\
unhtrema3 & 0.03 & 0 & 0 & 	0.134		& 	0.089	\\
unhtrema2 & 0.01 & 0 & 0 	& 	0.090		& 	0.131\\
unhtrema1 & 0 & 0 & 0	& 	0.061		& 	0.156 \\
     \hline
     Human description & 1.45 & 28 & 71 \\
     \hline
Baseline filtered & 1.49 & 33 & 71 \\
Baseline bartpodcasts & 1.49 & 25 & 75 & 	0.210		& 	0.208	\\
Baseline bartcnn & 0.99 & 10 & 35& 	0.272		& 	0.085		 \\
Baseline onemin & 0.93 & 5 & 30 & 	0.282		& 	0.087\\
Baseline textranksegments & 0.38 & 3 & 9	& 	0.165		& 	0.083	 \\
Baseline textranksentences & 0.23 & 1 & 4& 	0.162		& 	0.065 \\
     \hline
    \end{tabular}
    \caption{Overview of manual assessment results from submitted summarization experiments. The aggregate EGFB score is computed by assigning E=4, G=2, F=1, B=0. ROUGE scores are computed against the original creator provided descriptions of each episode.
    }
    \label{tab:manualsummarisation-results}
\end{table*}

\section{Task evolution for Year 2}
For 2021, we have the ambition to encourage participants to make use of the audio data in addition to the transcripts, but we do not wish to change the overall task formulation. 

We intend to continue the segment retrieval task with some modifications. In 2020, the segment retrieval output was restricted to two-minute segments at fixed starting points over the episode: in 2021, we will consider freely selected jump-in points in the episode, to allow for more precise segment results. We will add new topic types to the three used this year, including types that are likely to be better addressed if the audio signal is taken into consideration.  

We will specify the use case which the summarization task is intended to address in greater detail, with the target notion being an Audio Trailer, i.e. the output of the task should be a short highlight clip from the podcast episode in question. In practice, this means that the clip is not required to provide a representation of the entire content but an indicative segment which will inspire the listener to listen to the entire episode. The details of this specification will be formulated to make assessment transparent and reproducible. 

\section*{Acknowledgments}
Building and distributing the dataset and task definitions for the TREC 2020 Podcast Track was a highly collaborative effort, and we have many people to thank for their valuable contributions: Mounia Lalmas-Roelleke, Claire Burke, Zachary Piccolomini, Derek Tang, David Riordan, Fallon Chen, Lex Beattie, Daren Gill, Laura Pezzini, Jenni Lee, Alexandra Wei, Johannes Vuorensola, Nir Zicherman, Max Cutler, Julia Kaplan, Ching-Wei Chen, Brian Brost, Till Hoffmann, Nagarjuna Kumarappan, Maria Dominguez, Laurence Pascall, Md Iftekhar Tanveer, Rezvaneh Rezapour, Hamed Bonab and Jen McFadden.

\pagebreak

\bibliographystyle{unsrtnat}
\bibliography{JonesEtAl-Podcast-Overview-TREC} 
\end{multicols}
\end{document}